# Ultra-dense Radio Access Networks for Smart Cities: Cloud-RAN, Fog-RAN and "cell-free" Massive MIMO

*(Invited Paper)*

Alister Burr, Manijeh Bashar and Dick Maryopi
Dept of Electronic Engineering
University of York, U.K
alister.burr@york.ac.uk

*Abstract*—**In this paper we discuss the requirements for a radio access network architecture for ultra-dense networks for "smart city" applications, and show that coordination is required between access points to overcome the effects of interference. We propose a new paradigm, Fog Massive MIMO, based on a combination of the "cell-free" massive MIMO concept and the Fog Radio Access Network (F-RAN). In particular we analyze the potential benefit of improved coordination between APs over different coordination ranges.**

## I. INTRODUCTION

A wide range of new applications for wireless networks are currently appearing to enhance the environment of our "smart cities" [1]. These include wireless devices embedded in our homes and our industry, our vehicles and transport networks, our energy supply networks, and other infrastructure networks. These embedded devices, in addition to the plethora or wireless terminals that we all use, mean that the density of wireless devices in our cities will soon increase to several orders of magnitude more than the density of the human population.

A wireless network that can evolve to serve these applications must therefore become an *ultra-dense network* (UDN). Such networks have been defined as networks in which the density of access points (APs) may reach or exceed that of user terminals, where inter-AP distance is a few metres, or where the impact of interference is such that the capacity scaling of the conventional cellular paradigm begins to fail [2].

This latter issue in particular is the feature of UDNs that calls for a new approach to access network technology. Conventional cellular networks rely on the assumption that path losses for intercellular interference and for the signal path scale in the same way as cell sizes diminish. Until now this has allowed the capacity-density of a cellular system to increase without limit by increasing the AP density. However as the inter-AP distance becomes comparable with the scale of buildings and other features of the radio environment, this relationship may no longer hold, as interferers may now be in line of sight (LoS), and hence capacity remains limited by interference as network density increases.

It has been known for many years now that intercellular interference can effectively be eliminated by enabling coordination between APs, in an approach now often known as *network MIMO* [3,4]. Here the AP antennas can be regarded as the elements of a very large distributed antenna array in a multiuser MIMO (MU-MIMO) system encompassing the whole access network. This insight led also to the concept of *coordinated multipoint* (CoMP) [5] in 3GPP standards. More recently this approach has led to the concept of *cloud-RAN* (C-RAN) [6,7], in which the radio access network (RAN) over an area of possible tens of $km^2$ is treated as a distributed antenna system (DAS) in which AP antennas are connected to a large central *baseband processing unit* (BBU), at which all physical and higher layer functions of a base station will be performed, via "fronthaul" connections carrying quantized signals. The objective here is also to enable greatly reduced complexity at the AP locations (since they then contain only antennas, RF hardware and digital/analog conversion), but it clearly has the effect of reducing the RAN over this entire area to a single MU-MIMO system.

However for the much wider range of applications envisaged for "smart cities", many of which require limited latency, C-RAN has disadvantages, since concentrating processing to a remote location can result in significantly increased delay, especially if the services provided are essentially local. The computation carried out in the remote BBU will also be very complex, which may give rise to additional delay. In addition, the capacity of the fronthaul network, since it carries quantized signals rather than data, must be many times the total user data rate. More recently still, therefore, and especially in view of the requirement for ultra-reliable low latency communications (URLLC) in the fifth-generation (5G) mobile communication standards [8], there have been proposals to move the processing back from the "cloud" towards the network edge (i.e. closer to APs) – a location sometimes referred to as the "fog" (as opposed to the "cloud").

This has led to a new paradigm called *Fog-RAN* (F-RAN), in which communications, storage and computing functions are moved either into or closer to the APs at the network edge. The term was apparently first coined at the Next Generation Mobile Networks (NGMN) Forum in June 2014, and has given rise to a range of research (e.g. [9-12]), discussing variations on the proposed network architecture and focusing on different network functions.

While F-RAN has broader objectives than the implementation of AP coordination, it does allow us to consider how close to the network edge it is either necessary or desirable to implement such coordination. While the C-RAN approach enables full coordination of large numbers of APs over a very wide area, in fact the interactions between APs arise much more locally, at distances over which significant interference can occur between user signals. This suggests that coordination and joint processing can be carried out by an entity which has direct connections with only a few APs. Assessing the trade-offs between the coordination area of such an entity and the overall network performance is the main objective of this paper.

An issue that arises with any coordinating entity is the edge effect. If the APs are partitioned between several coordinating entities, there will inevitably be edges of the coordination areas where adjacent APs are not coordinated, and hence these parts of the network tend to have poorer performance. It is the minimization of such effects that tends to lead to very large coordination areas. Hence an essential part of our vision of F-RAN is that the coverage of adjacent entities should overlap, so that some APs are coordinated by more than one. Ultimately the objective is *user-centric* processing, where each user's signals are processed via the most appropriate subset of APs, and as close to that user as possible.

Another new paradigm for next generation access networks is of course *Massive MIMO* (MaMIMO), which again relies upon large numbers of coordinated antennas serving multiple users based on an MU-MIMO approach. In "classical" MaMIMO, however, the antennas are collocated at a base station in the centre of a cell, where the associated baseband processing also takes place. The very large number of antennas results in a phenomenon known as channel hardening, which effectively eliminates multipath fading. However the users located far from the base station are still relatively disadvantaged as a result of the path loss. More recently "cell-free" MaMIMO has been proposed [13,14]. Here the antennas are distributed across the cell, giving much improved service for the former cell-edge users, hence motivating the term "cell-free". (This could also be referred to as *distributed MaMIMO*, and we will use the abbreviation D-MaMIMO here). Signals are then conveyed to a *central processing unit* (CPU), where baseband and other processing is carried out. We have previously pointed out the conceptual similarity of this architecture to C-RAN [15], and the usefulness of the model introduced in [13] to evaluate the physical layer performance of such an architecture. In this paper we will adapt the approach for the analysis of the F-RAN architecture, leading to what we may call "Fog Massive MIMO" (F-MaMIMO). Note however that it has been shown [16] that one cannot rely on the "channel hardening" effect in "cell-free" massive MIMO except under two assumptions: that there are multiple antennas per AP, and that the path loss exponent is low. Both these conditions naturally apply to our proposed F-MaMIMO paradigm.

The remainder of this paper is structured as follows. In the next section we describe the F-MaMIMO system in more detail, while in Section III we describe our analysis methodology, in Section IV we give some numerical results, and in Section V we conclude the paper.

## II. Architecture Description and System Model

The proposed architecture is illustrated in Fig. 1. which shows a portion of the network. The edge-located CPUs are referred to as *edge processing units* (EPU); each coordinates (possibly jointly with adjacent EPUs) the APs within a *coordination region* (shown by the circles in Fig. 1), of radius $r_{coor}$. Note that the coordination regions of adjacent EPUs typically overlap, so that some APs are coordinated by more than one EPU. There is additionally an exclusive *service region* for each EPU (shown by the dotted hexagons); the data of users within this region is fully decoded at that EPU. The distance between centres of coordination regions is $d_{EPU}$, which here is assumed to be 1 km. Fronthaul connections are provided between each AP and all EPUs within whose service region the AP lies. (In this paper we do not discuss the technology by which these are provided, or any limitations it may have: we assume fronthaul connections are error-free and of unlimited bandwidth).

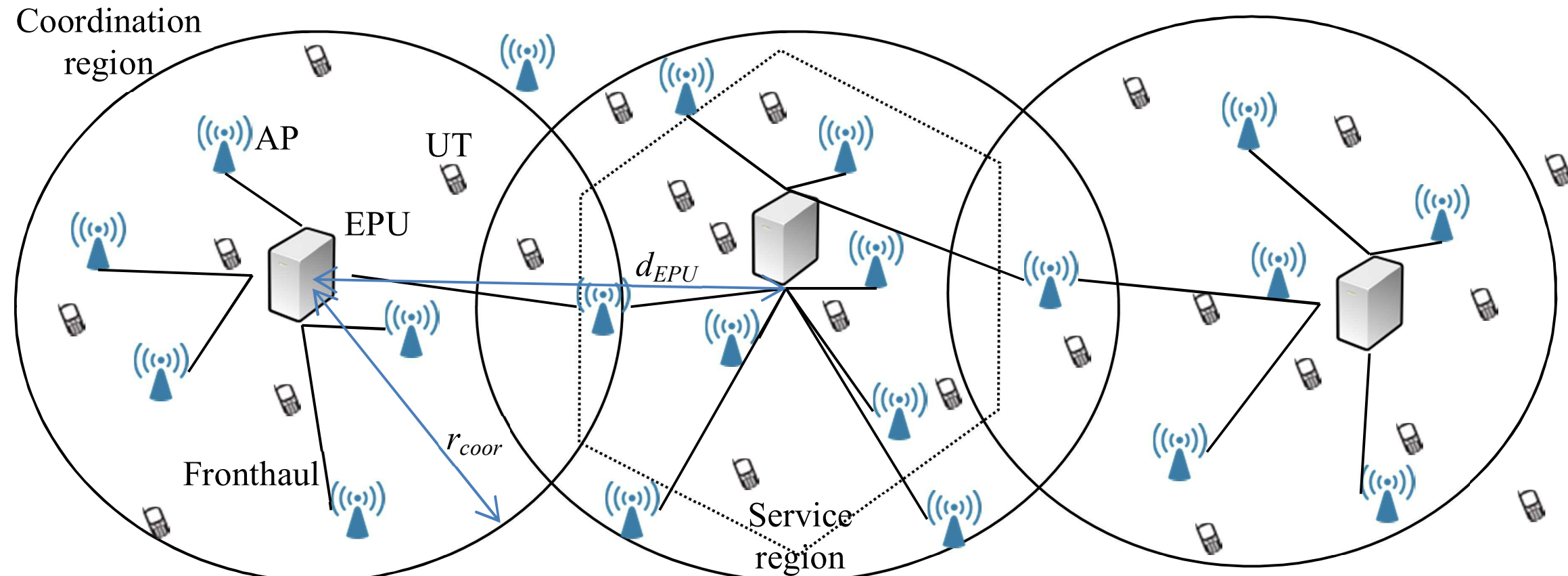

Fig. 1. F-MaMIMO architecture, showing edge processing units (EPU), with their coordination regions, access points (AP) with fronthaul connections to EPUs, and user terminals (UT)

We assume here that user terminals (UTs) have only a single antenna; APs may have $N_r$ antennas. Both UTs and APs are assumed to be uniformly distributed with densities $\rho_u$ and $\rho_A$ respectively. We assume that the flat-fading radio channel between antenna $n$ of AP $m$ and UT $k$ has gain (in general complex) $g_{mnk}$, given by:

$$g_{mnk} = h_{mnk}\sqrt{\beta_{mk}} \qquad (1)$$

in which $h_{mnk} \sim \mathcal{CN}(0,1)$ represents Rayleigh fading, and the real value $\beta_{mk}$ includes path loss, using a three-slope distance law:

$$\beta_{mn} = \begin{cases} 1 & d_{mn} < d_0 \\ \left(\dfrac{d_{mn}}{d_0}\right)^{-\gamma_0} & d_0 \le d_{mn} < d_1 \\ \left(\dfrac{d_1}{d_0}\right)^{-\gamma_0}\left(\dfrac{d_{mn}}{d_1}\right)^{-\gamma_1} & d_1 \le d_{mn} \end{cases} \qquad (2)$$

Here $\gamma_0 = 2$ and $\gamma_1 = 3.5$, $d_0 = 10$ m, $d_1 = 50$ m, as in [14].

We assume the same transmission format commonly assumed for MaMIMO, except that here we discuss a single-carrier system on a flat fading channel: we do not consider the application of OFDM. (Of course we may regard this channel as a single subcarrier of an OFDM system). We assume that the Rayleigh fading is constant over a coherence time of at least $\tau_c$, which limits the maximum packet length to this value, measured in channel uses (and thus taking an integer value). A period of $\tau_p$ channel uses is reserved for a pilot transmission (on the uplink), leaving a period $\tau_d = \tau_c - \tau_p$ for data transmission. UT $k$ transmits the length $\tau_p$ pilot sequence $\mathbf{p}_k$, which is received by the set of antennas at all APs connected to the serving EPU and used to calculate the channel vector $\mathbf{g}_k = \{\mathbf{g}_{1k} \;\cdots\; \mathbf{g}_{mk} \;\cdots\; \mathbf{g}_{Mk}\}^T$ corresponding to this UT, where $\mathbf{g}_{mk}$ is the $(N_r \times 1)$ vector of channels from the $k$th UT to the $N_r$ antennas of the $m$th AP. This channel vector is also used on the downlink, with the assumption of channel reciprocity – however we do not discuss the downlink in this paper.

We will assume here that all UTs in the coordination region of a given EPU use a set of orthogonal pilots: it follows that the minimum pilot length is $\tau_p \ge K_{coor}$, where the expected number of UTs in the coordination region, $K_{coor} = \pi r_{coor}^2 \rho_u$. Random pilots are assigned to users outside this region. Note that the service area is given by $\frac{1}{2} \times \frac{d_{EPU}}{\sqrt{3}} \times \frac{d_{EPU}}{2} \times 6 = \frac{\sqrt{3}}{2} d_{EPU}^2$, and hence the expected number of UTs served by this EPU is $K_{serv} = \frac{\sqrt{3}}{2} \rho_u d_{EPU}^2$.

## III. Performance Analysis

In this section we estimate the performance of our F-MaMIMO system, and in particular how the area of the coordination region relative to the service region affects performance. Increasing the radius $r_{coor}$ of the coordination region means that a larger proportion of the signal power from the UTs served can be collected by the $M_{coor}$ APs which are coordinated, where $M_{coor} = \rho_A \pi r_{coor}^2$. In addition, the signals received from all UTs in this region are also coordinated, and therefore do not in principle cause interference, and hence uncoordinated interference also decreases as $r_{coor}$ increases. However this requires pilots of length $\tau_p = K_{coor}$ in order to serve only $K_{serv}$ UTs, which requires a longer pilot period $\tau_p$.

### A. Signal power

Considering first the potential signal power improvement, let the length $N_r$ vector of received signals at the $N_r$ antennas on the $m$th AP due to the $k$th UT be:

$$\mathbf{y}_{mk} = \mathbf{g}_{mk} x_k = \mathbf{h}_{mk} \sqrt{\beta_{mk}}\, x_k \tag{3}$$

The total signal power collected from UT $k$ is therefore:

$$\begin{aligned} S_k &= \sum_{m=1}^{M_{coor}} \left\| \mathbf{y}_{mk} \right\|^2 = \sum_{m=1}^{M_{coor}} \left\| \mathbf{h}_{mk} \right\|^2 \beta_{mk} \left| x_k \right|^2 \\ &\approx N_r\, \sigma_x^2 \sum_{m=1}^{M_{coor}} \beta_{mk} \end{aligned} \tag{4}$$

### B. Channel Estimation

Assuming a sufficient degree of channel hardening and favourable propagation (as discussed in section I above), the primary source of interference is due to pilot contamination. Hence we first consider channel estimation.

As mentioned above, we assume that orthogonal pilots are transmitted within the coordination region of one EPU. These can be described by a $\tau_p \times K_{serv}$ matrix $\mathbf{P}$ whose columns are the pilot sequences of the $K_{serv}$ users, such that $\mathbf{P}^H \mathbf{P} = \tau_p \mathbf{I}_{K_{serv}}$. We suppose that users outside this coordination region use pilot sequences which are not orthogonal to the columns of $\mathbf{P}$. We will define these as the columns of a $\tau_p \times K_{int}$ matrix $\mathbf{P}_{int}$, where $K_{int}$ denotes the number of significant interfers. (The number of interferers is of course in principle unlimited, but we include in $K_{int}$ those that are close enough to affect the channel estimation of the target EPU). We write $\mathbf{C}_{p,int} = \mathbf{P}^H \mathbf{P}_{int} / \tau_p$, and assume that:

$$\begin{aligned} &\mathbb{E}\left[ \mathbf{C}_{p,int} \right] = \mathbf{0}_{K_{serv} \times K_{int}} \\ &\mathbb{E}\left[ \left| \mathbf{C}_{p,int} \right|^2 \right] = \mathbf{1}_{K_{serv} \times K_{int}} / \tau_p \end{aligned} \tag{5}$$

The expectation here is taken over the randomly assigned pilot sequences. The received pilot signals at the $M_{coor}$ APs coordinated by the target EPU, including interfering pilots but neglecting noise (on the assumption of large interference to noise ratio), may be represented by the $M_{coor} \times \tau_p$ matrix:

$$\mathbf{Y}_p = \mathbf{G}\mathbf{P}^T + \mathbf{G}_{int} \mathbf{P}_{int}^T + \mathbf{Z} \tag{6}$$

where $\mathbf{G}$ denotes the $N_r M_{coor} \times K_{serv}$ channel matrix between the served users and the antennas of the coordinated APs, and $\mathbf{G}_{int}$ the $N_r M_{coor} \times K_{int}$ channel matrix from the interfering users. $\mathbf{Z}$ denotes uncorrelated Gaussian noise at receive antennas, each element having standard deviation $\sigma_n$. We may then estimate the channel matrix:

$$\begin{aligned} \hat{\mathbf{G}} &= \frac{1}{\tau_p} \mathbf{Y}_p \mathbf{P}^* \\ &= \frac{1}{\tau_p} \mathbf{G}\mathbf{P}^T \mathbf{P}^* + \frac{1}{\tau_p} \mathbf{G}_{int} \mathbf{P}_{int}^T \mathbf{P}^* + \frac{1}{\tau_p} \mathbf{Z}\mathbf{P}^* \\ &= \mathbf{G} + \tilde{\mathbf{G}} = \mathbf{G} + \mathbf{G}_{int} \mathbf{C}_{p,int}^T + \mathbf{C}_{p,z} \end{aligned} \tag{7}$$

### C. Data Detection

We then use this channel estimate to detect the data from the corresponding received signal matrix. In this case it is sufficient to consider a vector of signals from all UTs in one symbol period. Then the received signal vector:

$$\mathbf{y}_d = \mathbf{G}\mathbf{x} + \mathbf{G}_{int} \mathbf{x}_{int} + \mathbf{z} \tag{8}$$

We then form a (scaled) data symbol estimate:

$$\begin{aligned} \hat{\mathbf{x}} &= \hat{\mathbf{G}}^H \mathbf{y}_d = \hat{\mathbf{G}}^H \mathbf{G}\mathbf{x} + \hat{\mathbf{G}}^H \mathbf{G}_{int} \mathbf{x}_{int} + \hat{\mathbf{G}}^H \mathbf{z} \\ &= \mathbf{G}^H \mathbf{G}\mathbf{x} + \mathbf{C}_{p,int}^* \mathbf{G}_{int}^H \mathbf{G}\mathbf{x} + \mathbf{C}_{p,z}^H \mathbf{G}\mathbf{x} \\ &\quad + \mathbf{G}^H \mathbf{G}_{int} \mathbf{x}_{int} + \mathbf{C}_{p,int}^* \mathbf{G}_{int}^H \mathbf{G}_{int} \mathbf{x}_{int} \\ &\quad + \mathbf{C}_{p,z}^H \mathbf{G}_{int} \mathbf{x}_{int} + \hat{\mathbf{G}}^H \mathbf{z} \end{aligned} \tag{9}$$

We make the simplifying assumption that channel hardening applies, allowing us to replace the terms in this expression by their expectation over Rayleigh fading. In the second and fifth terms the columns of $\mathbf{G}$ and $\mathbf{G}_{int}$ are uncorrelated, and hence these terms disappear. We are then left with a signal term:

$$\begin{aligned} &\mathbb{E}\left[ \mathbf{G}^H \mathbf{G} \right] \mathbf{x} \\ &= \operatorname{diag}\left[ N_r \sum_{m=1}^{M_{coor}} \beta_{mk}, k = 1 \ldots K \right] \mathbf{x} \end{aligned}, \tag{10}$$

plus an interference term due to pilot contamination:

$$\begin{aligned} &\mathbf{C}_{p,int} \mathbb{E}\left[ \mathbf{G}_{int}^H \mathbf{G}_{\mathrm{int}} \right] \mathbf{x}_{int} \\ &= \mathbf{C}_{p,int} \operatorname{diag}\left[ N_r \sum_{m=1}^{M_{coor}} \beta_{mk'}, k' = 1 \ldots K_{int} \right] \mathbf{x}_{int} \end{aligned} \tag{11}$$

The remaining terms are noise, and it turns out that they are dominated by the term $\hat{\mathbf{G}}^H \mathbf{z} \approx \mathbf{G}^H \mathbf{z}$ (because the terms containing $\mathbf{C}_{p,z}^H$ have variance $\sigma_z^2 / \tau_p$ ).

Hence the signal to noise ratio for the $k$th user:

$$\begin{aligned} SNR_k &= \frac{\sigma_x^2 \left( N_r \sum_{m=1}^{M_{coor}} \beta_{mk} \right)^2}{\sigma_z^2 N_r \sum_{m=1}^{M_{coor}} \beta_{mk}} \\ &= \frac{\sigma_x^2}{\sigma_z^2} N_r \sum_{m=1}^{M_{coor}} \beta_{mk} \end{aligned} \tag{12}$$

The interference to noise ratio:

$$INR_k = \frac{\sigma_x^2 N_r}{\sigma_z^2 \tau_p} \frac{\sum_{k'=1}^{K_{int}} \left( \sum_{m=1}^{M_{coor}} \beta_{mk'} \right)^2}{\sum_{m=1}^{M_{coor}} \beta_{mk}} \qquad (13)$$

and the signal to interference-plus-noise ratio:

$$SINR_k = \frac{\left( \sum_{m=1}^{M_{coor}} \beta_{mk} \right)^2}{\frac{1}{\tau_p} \sum_{k'=1}^{K_{int}} \left( \sum_{m=1}^{M_{coor}} \beta_{mk'} \right)^2 + \frac{\sigma_z^2}{\sigma_x^2 N_r} \sum_{m=1}^{M_{coor}} \beta_{mk}} \qquad (14)$$

although in practice the noise term here (the second term in the denominator) can usually be neglected.

## IV. Numerical Results

### *A. Performance of CF-MaMIMO*

For comparison we first evaluate the performance of CF-MaMIMO, taking into account uncoordinated interference from outside the region covered by a CPU. The performance analysis is that described in section III, except that the coordination area is coincident with the service area: there is no overlap. The system parameters are based on those used in [14], in which 40 UTs and 100 APs are distributed in a service area 1 km square. We consider a denser system in which $\rho_u$ = 160/km$^2$ and $\rho_A$ = 400/km$^2$. Fig. 2 shows a scatter plot of user locations with their SIRs, assuming that the distance between CPUs is 1 km, showing that some users close to the edge of the service area are particularly disadvantaged. Fig. 3 shows the cumulative distribution function (CDF) of these SIRs (black plot), showing however that in this case the considerable majority of users achieve an adequate SIR.

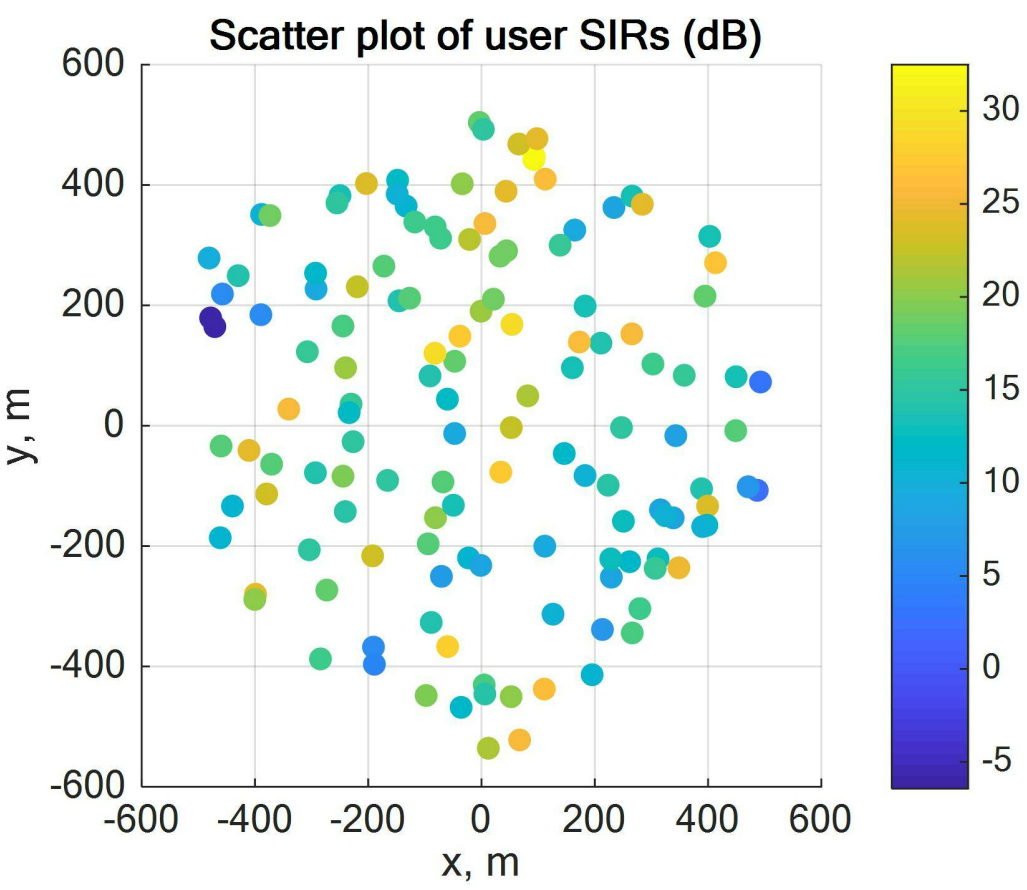

Fig. 2. Scatter plot of UTs served by one CPU, showing their SIRs

However if the size of the coverage area is reduced, so as to bring the CPU closer to the users as discussed in section I, the resulting CDF is given by the red line in Fig. 3, showing that now a high proportion of users have a negative SIR. This is because the edge effects now encompass a much higher proportion of users. This demonstrates the need for overlapping coordination areas, as discussed above.

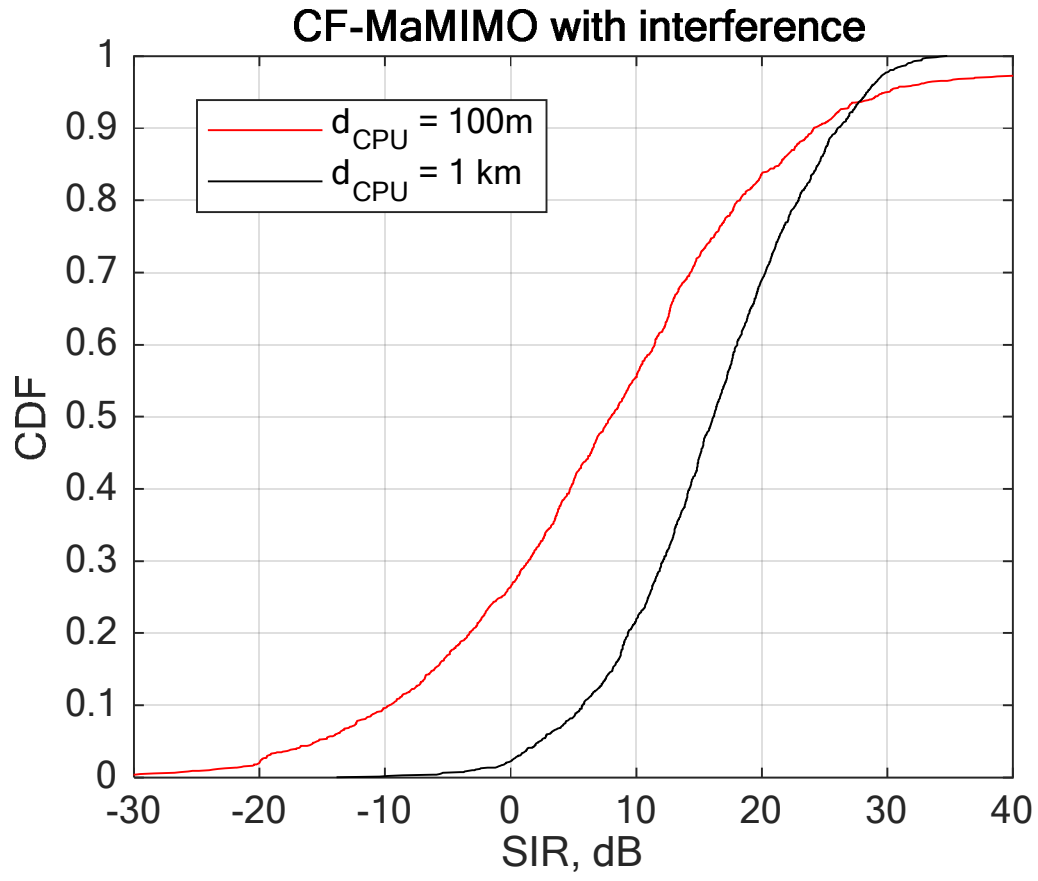

Fig. 3. Cumulative distribution function of SIRs for UTs served by one CPU, for different inter-CPU distances

We next evaluate our proposed F-MaMIMO architecture, with EPU spacing 100m, to determine the coordination range required. Fig. 4 shows the distribution of the total power received from each UT at all the coordinated APs. It shows that at the 95$^{th}$ centile a gain of around 7 dB is available, but there are diminishing returns beyond 100m. Note that 58m is the radius of the circle required to fully cover the service area of the EPU.

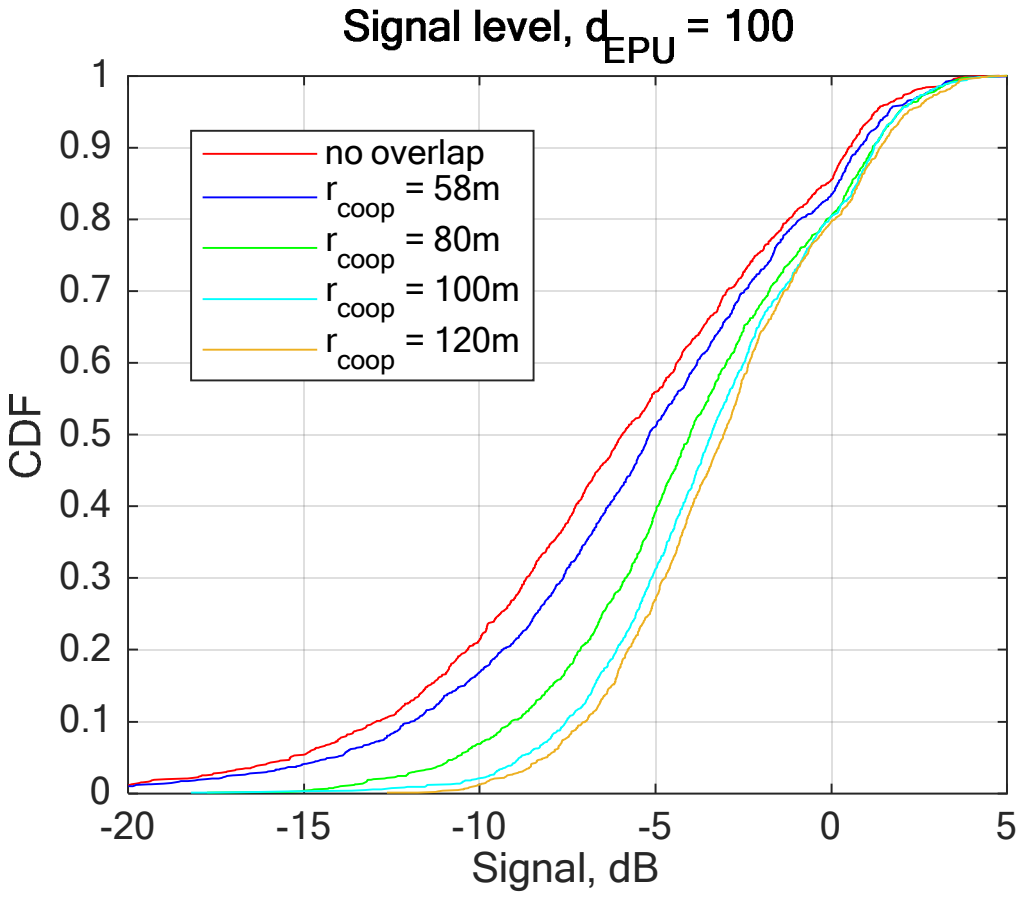

Fig. 4. CDF of signal power of UTs received by coordinated APs, for different coordination radii. The red plot assumes the coordination area is the service area only

Fig. 5 similarly shows the CDF of SIR experienced by UTs (neglecting noise, since in general ranges are small). The overlap allows a significant improvement in SIR: 12 dB and almost 30 dB for $r_{coor}$ = 80m and 120m respectively at the $95^{th}$ centile. Note that, in contrast to the D-MaMIMO case where 139 users are expected within the service area, and hence the pilot sequences must be at least this long, even with $r_{coor}$ = 120m only 8 users are expected within the coordination region, and hence very short pilot sequences are possible.

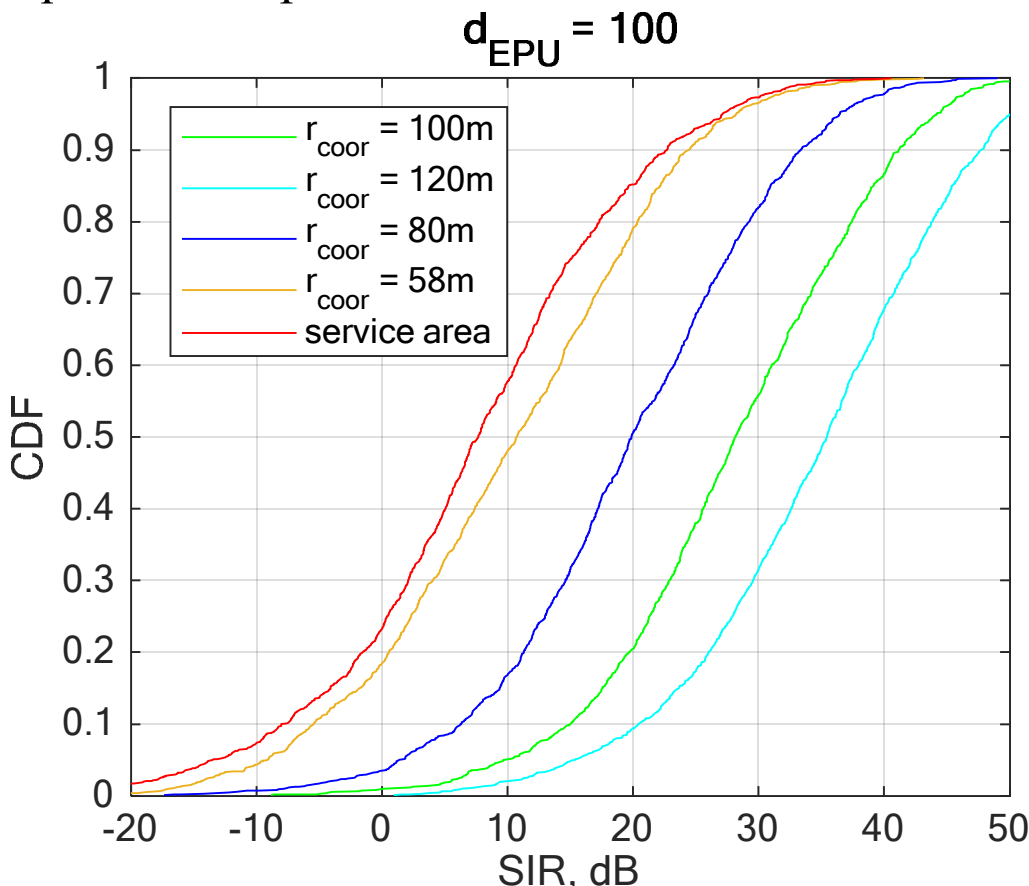

Fig. 5. CDF of SIR of UTs for different coordination radii. Red plot assumes coordination area is coverage area, with no overlap

## V. Conclusions

This paper has introduced a new paradigm for the Fog Radio Access Network based on the "cell-free" Massive MIMO concept, which we describe as F-MaMIMO. Rather than bringing signals from APs to a single central processing unit, they are distributed among a set of *edge processing units* (EPUs) with overlapping coverage, which coordinate the signals between neighbouring APs. We the effect of increasing the area over which APs are coordinated, and show that significant improvement in user SIR are possible, while still requiring only very short pilot sequences.

## Acknowledgment

The work described in this paper is supported in part by the European Commission under MSCA SPOTLIGHT, grant no. 722788.

## References

[1] Department for Business Innovation and Skills "Smart Cities: Background Paper" October 2013. Online: https://assets.publishing.service.gov.uk/government/uploads/system/uploads/attachment_data/file/246019/bis-13-1209-smart-cities-background-paper-digital.pdf

[2] Jialing Liu, Weimin Xiao, Chih-Lin I, Chenyang Yang, Anthony Soong "Ultra-Dense Networks (UDNs) for 5G" IEEE 5G Tech Focus: Volume 1, Number 1, March 2017. Online: https://5g.ieee.org/tech-focus/march-2017/ultra-dense-networks-udns-for-5g

[3] M. K. Karakayali, G. J. Foschini and R. A. Valenzuela, "Network coordination for spectrally efficient communications in cellular systems," in *IEEE Wireless Communications*, vol. 13, no. 4, pp. 56-61, Aug. 2006.

[4] D. Gesbert, S. Hanly, H. Huang, S. Shamai Shitz, O. Simeone and W. Yu, "Multi-Cell MIMO Cooperative Networks: A New Look at Interference," in *IEEE Journal on Selected Areas in Communications*, vol. 28, no. 9, pp. 1380-1408, December 2010

[5] R. Irmer *et al.*, "Coordinated multipoint: Concepts, performance, and field trial results," in *IEEE Communications Magazine*, vol. 49, no. 2, pp. 102-111, February 2011

[6] “C-RAN: the road towards green RAN,” China Mobile Research Institute, Beijing, China, Oct. 2011, Tech. Rep.

[7] A. Checko *et al.*, "Cloud RAN for Mobile Networks—A Technology Overview," in *IEEE Communications Surveys & Tutorials*, vol. 17, no. 1, pp. 405-426, Firstquarter 2015

[8] Chih-Ping Li, Jing Jiang, W. Chen, Tingfang Ji and J. Smee, "5G ultra-reliable and low-latency systems design," *2017 European Conference on Networks and Communications (EuCNC)*, Oulu, 2017

[9] Y. J. Ku *et al.*, "5G Radio Access Network Design with the Fog Paradigm: Confluence of Communications and Computing," in *IEEE Communications Magazine*, vol. 55, no. 4, pp. 46-52, April 2017

[10] S. H. Park, O. Simeone and S. Shamai Shitz, "Joint Optimization of Cloud and Edge Processing for Fog Radio Access Networks," in *IEEE Transactions on Wireless Communications*, vol. 15, no. 11, pp. 7621-7632, Nov. 2016.

[11] Y. Shi, J. Zhang, K. B. Letaief, B. Bai and W. Chen, "Large-scale convex optimization for ultra-dense cloud-RAN," in *IEEE Wireless Communications*, vol. 22, no. 3, pp. 84-91, June 2015

[12] M. Peng and K. Zhang, "Recent Advances in Fog Radio Access Networks: Performance Analysis and Radio Resource Allocation," in *IEEE Access*, vol. 4, pp. 5003-5009, 2016.

[13] HQ Ngo, A Ashikhmin, H Yang, EG Larsson, TL Marzetta. "Cell-free Massive MIMO: Uniformly great service for everyone". In *IEEE International Workshop on Signal Processing Advances in Wireless Communications* (SPAWC) 2015

[14] H. Q. Ngo, A. Ashikhmin, H. Yang, E. G. Larsson and T. L. Marzetta, "Cell-Free Massive MIMO Versus Small Cells," in *IEEE Transactions on Wireless Communications*, vol. 16, no. 3, pp. 1834-1850, March 2017.

[15] A. G. Burr, M. Bashar and D. Maryopi "Cooperative Access Networks: Optimum Fronthaul Quantization in Distributed Massive MIMO and Cloud RAN" *IEEE Vehicular Technology Conference* (VTC-S), Porto, 2018

[16] Z. Chen and E. Bjoernson, "Can We Rely on Channel Hardening in Cell-Free Massive MIMO?," *2017 IEEE Globecom Workshops (GC Wkshps)*, Singapore, 2017, pp. 1-6.